\pdfoutput=1
\documentclass[aps,manuscript,twocolumn, prl, reprint,english]{revtex4-1}
\usepackage[latin9]{inputenc}
\usepackage{graphicx}

\makeatletter

\@ifundefined{textcolor}{}
{%
 \definecolor{BLACK}{gray}{0}
 \definecolor{WHITE}{gray}{1}
 \definecolor{RED}{rgb}{1,0,0}
 \definecolor{GREEN}{rgb}{0,1,0}
 \definecolor{BLUE}{rgb}{0,0,1}
 \definecolor{CYAN}{cmyk}{1,0,0,0}
 \definecolor{MAGENTA}{cmyk}{0,1,0,0}
 \definecolor{YELLOW}{cmyk}{0,0,1,0}
}

\usepackage{epstopdf}
\usepackage{subfigure}

\usepackage[english]{babel}
\usepackage[T1]{fontenc}

\makeatother

\begin{document}

\title{Magnetically tuned, robust and efficient filtering system \\
for spatially multimode quantum memory in warm atomic vapors}

\author{Micha\l{} D\k{a}browski, Rados\l{}aw Chrapkiewicz$^{*}$, and
Wojciech Wasilewski}
\email{radekch@fuw.edu.pl}
\begin{abstract}
Warm atomic vapor quantum memories are simple and robust, yet suffer
from a number of parasitic processes which produce excess noise. For
operating in a single-photon regime precise filtering of the output
light is essential. Here we report a combination of magnetically tuned
absorption and Faraday filters, both light-direction-insensitive,
which stop the driving lasers and attenuate spurious fluorescence
and four-wave mixing while transmitting narrowband Stokes and anti-Stokes
photons generated in write-in and readout processes. We characterize
both filters with respect to adjustable working parameters. We demonstrate
a significant increase in the signal to noise ratio upon applying
the filters seen qualitatively in measurements of correlation between
the Raman-scattered photons.
\end{abstract}
\maketitle

\section{Introduction}

Quantum memories are the essential building-blocks in a number of
quantum information processing protocols \cite{Bussieres2013}. Examples
include building a quantum network connecting space-separated atomic
memories via optical fibers \cite{Kimble2008}, quantum computation
\cite{Lee2011a,Bustard2013}, quantum communication \cite{Kuzmich2003},
entanglement \cite{Lee2011} and multiple-photon generation on demand
\cite{Nunn2013}. 

A number of different implementations of quantum memory have been
realized so far including electromagnetically induced transparency
\cite{Veissier2013}, Raman scattering \cite{Reim2010}, atomic frequency
comb \cite{Hedges2010,Clausen2011}, gradient echo memory \cite{Hosseini2009}
and coherent population oscillations \cite{DeAlmeida2015}. Among
the most popular media for storage of the quantum information are
cold alkaline atoms \cite{Radnaev2010,Choi2010,Bao2012,Stack2015},
room-temperature atomic vapors in cells \cite{VanderWal2003,Reim2010,Hosseini2011,Chrapkiewicz2012,Bashkansky2012}
and in hollow-core fibers \cite{Sprague2013,Sprague2014}, molecular
gases \cite{Bustard2013,Bustard2015}, NV-centres \cite{Maurer2012}
and optical phonons \cite{Lee2011} in diamonds, and rare-earth-doped
solids \cite{DeRiedmatten2008,Saglamyurek2011}. 

In this work we focus on probably the most widespread and renowned
implementation based on warm alkali atom vapor (rubidium-87) contained
in a glass cell together with noble buffer gas (krypton) to reduce
the effects of thermal motion \cite{Parniak2013,Chrapkiewicz2014}.
Photons are interfaced with atomic collective excitations, spin-waves,
via off-resonant Raman transitions. We operate on memory in spatially
multimode regime using large diameter driving beams \cite{Koodynski2012b}.
In general multimode operation is natural and readily achievable in
a number of spontaneous and parametric processes \cite{Dabrowski2014a,Parniak2015,Parniak2015a}
such as spontaneous parametric down-conversion in non-linear crystals
\cite{Chrapkiewicz2010a} and instantaneous four-wave mixing in warm
atomic vapors \cite{Boyer2008,Boyer2008b}. In our application spatially
multimode operational mode can be applied to store transversely varying
information, such as images, in quantum memories \cite{Surmacz2008,Chrapkiewicz2012,Glorieux2012}.
In our setup typically over one hundred distinguishable spatial modes
of Raman light are coupled with independent spin-wave modes \cite{Chrapkiewicz2015}. 

\begin{figure}[b]
\includegraphics[width=1\columnwidth]{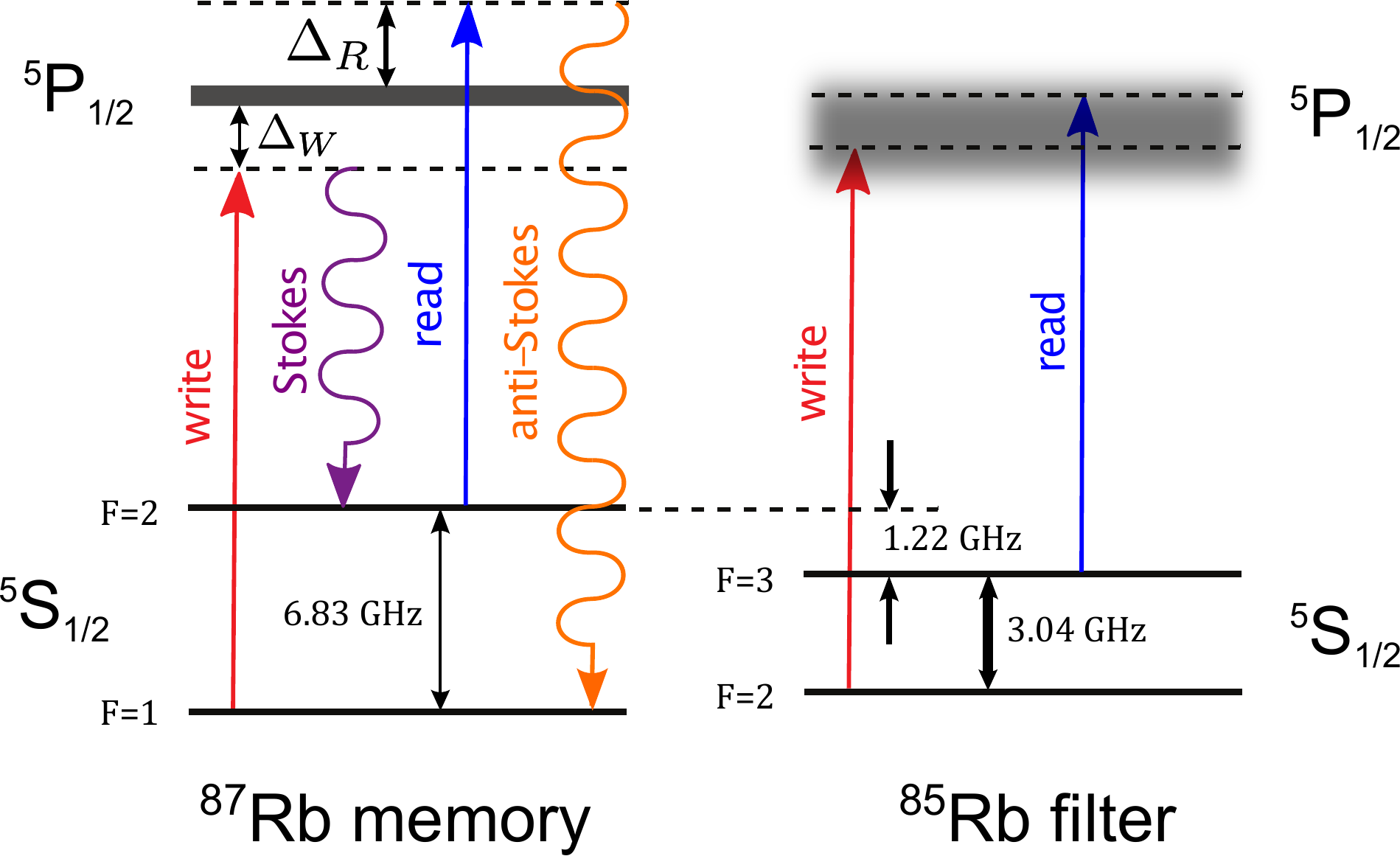} \centering
\protect\caption{Driving lasers and photons frequencies against rubidium-87 and rubidium-85
energy level structure. Write and read lasers are virtually resonant
with Doppler- and magnetically broadened $^{85}$Rb isotope transitions
in absorption filter as opposed to scattered photons. }
\label{fig:Rb-Levels} 
\end{figure}

Spontaneous Raman scattering in warm and high optical depth ensembles is most efficient at the detuning of the order of 1 GHz and requires significant drive laser energy.
Thus filtering out of the driving beams requires very large attenuation,
around $10^{11}$. In our experiments first we use crystal polarizers,
as the Raman photons are polarized orthogonally to the driving light.
Next a rubidium-85 absorption filter \cite{Heifetz2004,Stack2015}
of magnetically broadened absorption line width is used to stop any
remaining leakage. The second type of noise is an omnidirectional
spontaneous fluorescence directly from the excited state, observed
in a number of experiments \cite{Eisaman2004a,Manz2007,Bashkansky2012}
and discussed theoretically \cite{Shen1974,Rousseau1975,Raymer1977,Childress2005}.
This fluorescence is enhanced by collisions with buffer gas \cite{Eisaman2004a,Manz2007,Bustard2013}.
Our experiments also indicate presence of broadband and omnidirectional
noise light of an intensity proportional both to the concentration
of Rubidium and the buffer gas, likely due to the scattering of laser
light by the long-lived Rb-buffer gas molecules \cite{Hedges1972}. 

To filter out this light, a narrow bandpass filter transmitting the
Raman light only is necessary. Typically a Fabry-Perot cavity is used
to this end \cite{Kuzmich2003,Manz2007,Bashkansky2012}. However,
it is unsuitable for our purposes, because it cannot transmit hundreds
of spatial modes at the same frequency. Therefore we use a Faraday
filter \cite{Zielinska2012a,Zielinska2014} with rubidium-87 vapor
rotator as a second stage. By properly adjusting the temperature and
magnetic field in both filters we are able to attenuate the driving
beams by a factor of $10^{11}$ and reach a transmission of 65\% for
the Stokes and 45\% for the anti-Stokes light.

We demonstrate the immense effect of combined filtering on the statistical
properties of the registered light. The first filter virtually stops
the driving laser leakage. However, only due to the application of
the second filter intensity correlations between the Stokes and anti-Stokes
scattered light can be seen down to a single-photon level. The approach
we demonstrate is a robust alternative to filtering with the use of
interference cavities. It is insensitive to direction and precise
alignment although its tuning is possible just through adjusting the
magnetic field and temperature in both filters.

\section{Memory setup}

\begin{figure}[t]
\includegraphics[width=1\columnwidth]{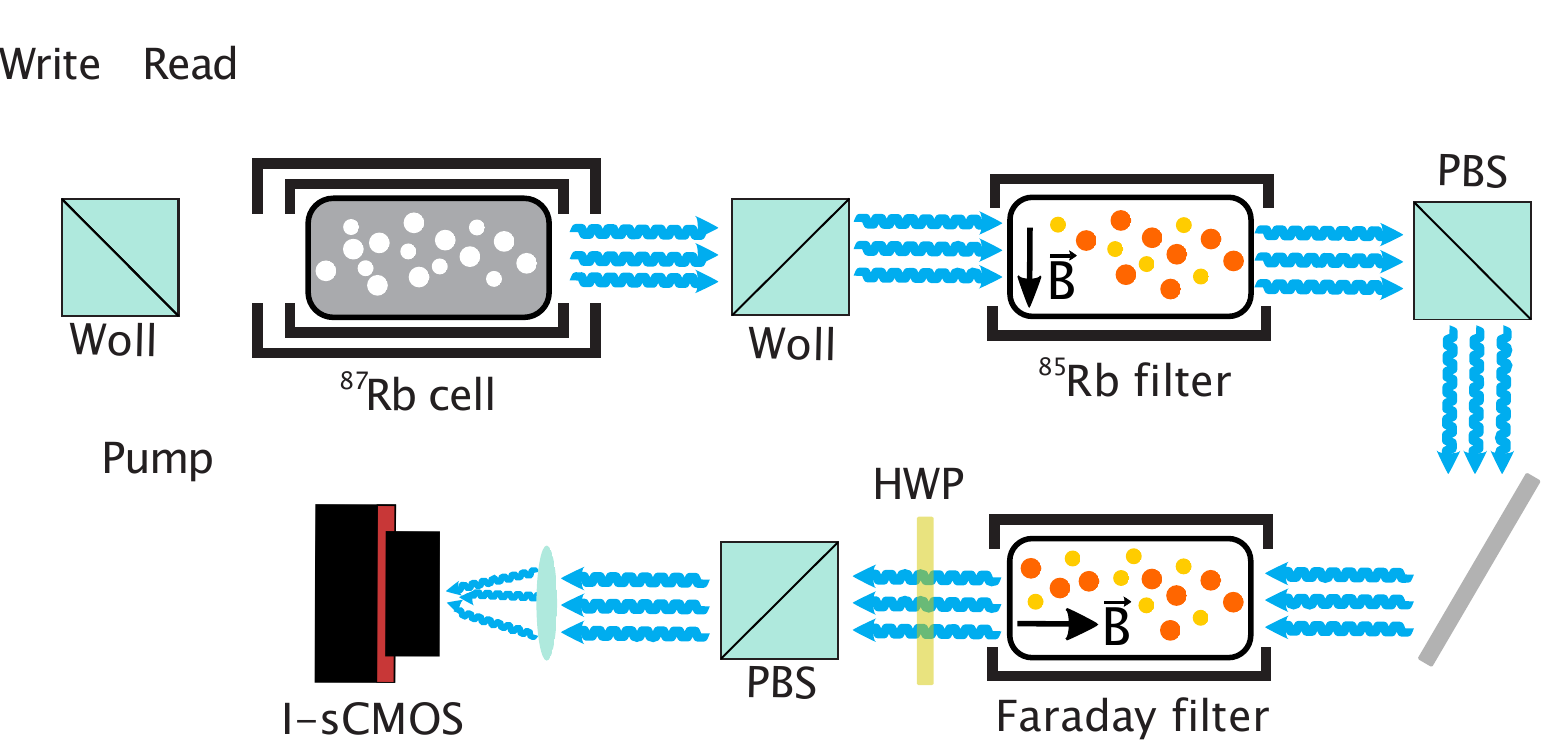} 
\centering
\protect\caption{Experimental setup of quantum memory and filters system. From top-left
corner: $^{87}$Rb atomic memory cell inside magnetic shielding, magnetically
broadened $^{85}$Rb absorption filter, Faraday filter and a low-noise,
single-photon-sensitive intensified sCMOS camera (I-sCMOS) as a detector.
The blue squares stand for polarizing beam splitter cubes (PBS) or
Wollaston polarizers (Woll). The directions of the magnetic field
are marked. }
\label{fig:setup} 
\end{figure}

In our experiments we start with an ensemble of $^{87}$Rb atoms pumped
to F=1 state. We write to the memory by driving spontaneous Stokes
transition, which produces pairs of collective excitations to the
F=2 state, i.e. spin-waves and scattered Stokes photons. After adjustable
storage time the spin-wave can be converted to anti-Stokes photons
by sending read laser pulse \cite{Kuzmich2003,Chrapkiewicz2012,Dabrowski2014}.
Both drive lasers work near D1-line transitions. The detunings of
write and read lasers from resonances $F=1\rightarrow F'=1$ and $F=2\rightarrow F'=2$
equal $\Delta_{W}=1$ GHz and $\Delta_{R}=800$ MHz respectively as
indicated in Fig.\ref{fig:Rb-Levels}.

The simplified version of the experimental setup is depicted in Fig.
\ref{fig:setup}. We utilize three external cavity diode lasers (ECDL)
for optical pumping and Raman scattering in the $^{87}$Rb cell. The
Raman scattered light has a polarization orthogonal to the driving
lasers, so we use Wollaston prisms to pre-filter the Raman-scattered
light. Due to the finite extinction of the Wollaston prisms ($10^{5}$:1)
a small portion of laser light leaks through together with the Raman-scattered
light. Usage of the $^{85}$Rb leads to a high absorption coefficient
of the driving lasers whereas transmission of Raman-scattered light
remains at the level of 80\%. Properties of the atomic absorption
filter are independent of the transmitting light properties, e.g.
direction, spatial mode profile or polarization, in contrast to the
properties of the Faraday or Voigt filters. Finally, the broadband
light coming from the resonant fluorescence is filtered out with the
use of the Faraday filter. This filter additionally diminishes the
residual leakage of driving beams.

The photons are detected with a high-resolution sCMOS camera with
an image intensifier (I-sCMOS) \cite{Chrapkiewicz2014e,Jachura2015,Jachura2015a,Chrapkiewicz2015f}.
This enables us to work in a single-photon-level regime and to measure
photon statistics in a similar manner as with\textbf{ }multiplexed
on-off detectors \cite{Chrapkiewicz2014f,Chrapkiewicz2014e,Jachura2015a,Chrapkiewicz2015f}.
Further details of the experimental setup and the full operational
scheme of our quantum memory protocol are described in \cite{Chrapkiewicz2012,Dabrowski2014}. 

\begin{figure}[t]
 \includegraphics[width=1\columnwidth]{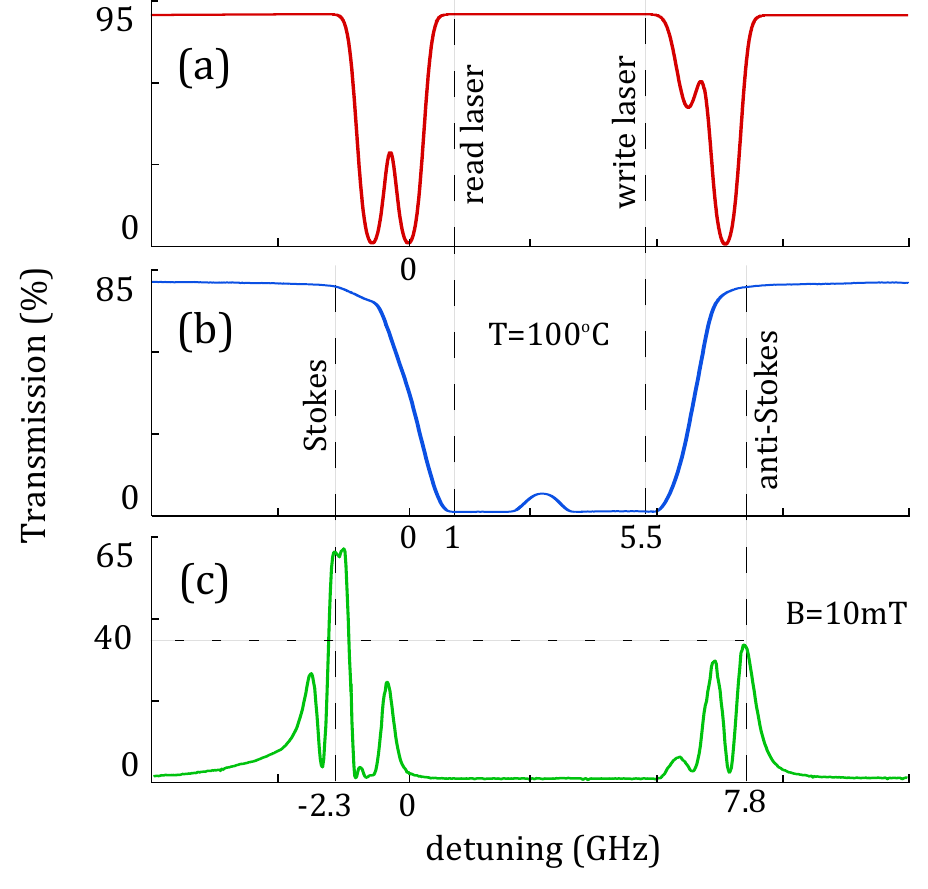}
 \centering
\protect\caption{Transmission spectra of (a) the atomic quantum memory cell containing
$^{87}$Rb isotope, (b) $^{85}$Rb absorption filter at $T=100^{\circ}$ C and (c) $^{87}$Rb Faraday filter for the magnetic field amplitude
$B=10^{-3}$ T and temperature $T=102^{o}$C. The frequency of the
laser beams and scattered Stokes and anti-Stokes light are marked
by detunings measured from $F=2\rightarrow F'=2$ transition on D1-line
in $^{87}$Rb.}
\label{fig:sample_transmissions} 
\end{figure}

Absorption and transmission spectra of the three rubidium cells used
in the experiment are presented in Fig. \ref{fig:sample_transmissions}
along with the positions of the laser beams frequencies, as well as
Stokes and anti-Stokes scattered light. Detuning of about 1 GHz for
both write-in and readout beams is a compromise between the high efficiency
of Raman scattering and the low absorption coefficient inside the
filtering cells.

\section{Filtering system - details}

\subsection{Absorption filter}

\begin{figure}[t]
\includegraphics[width=1\columnwidth]{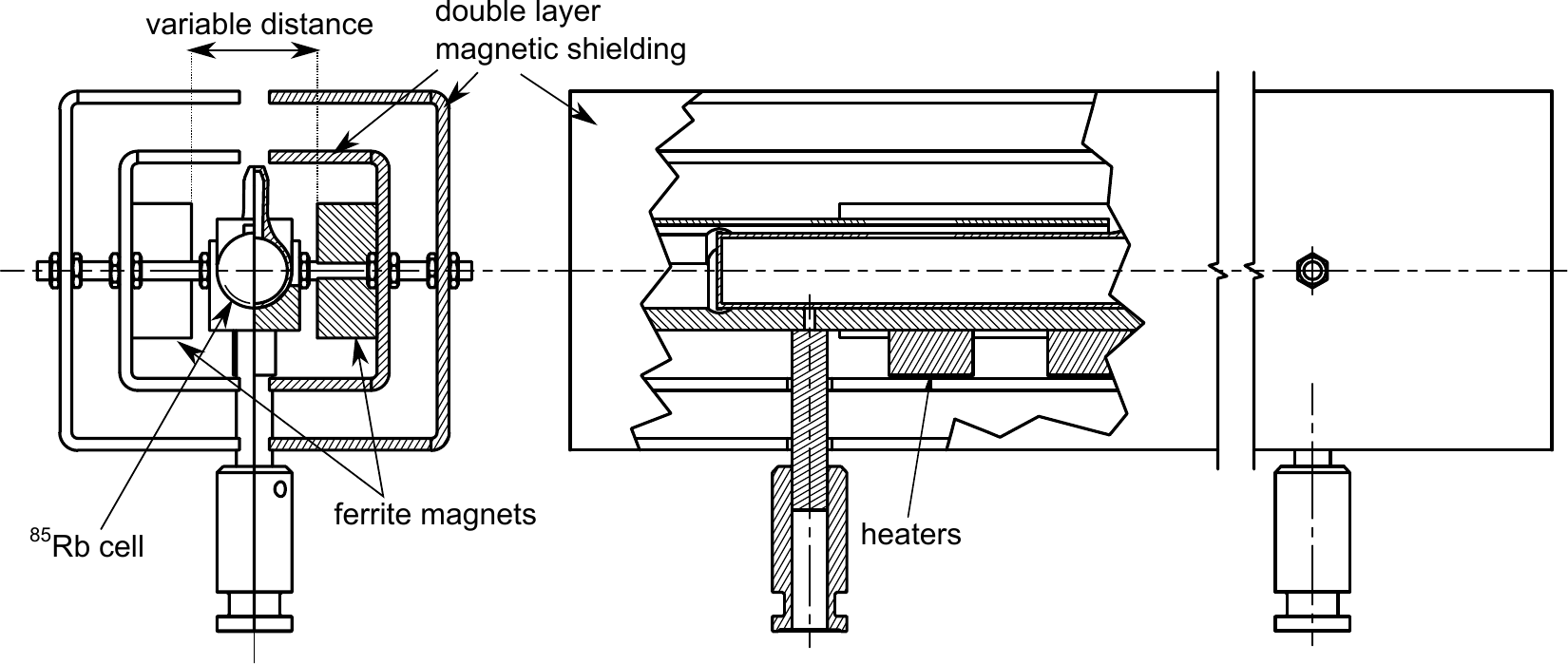}
\centering
\protect\caption{Technical drawing of the magnetically tuned absorption filter based
on $^{85}$Rb isotope. Broadening of filter lines is achieved by changing
a variable distance between ferrite magnets. Filter is embedded in
a two-layer magnetic shielding protecting main quantum memory cell
from the corrupting stray magnetic field. \label{fig:TechnicalDrawingFilter}}
\end{figure}

\begin{figure}[t]
\includegraphics[width=1\columnwidth]{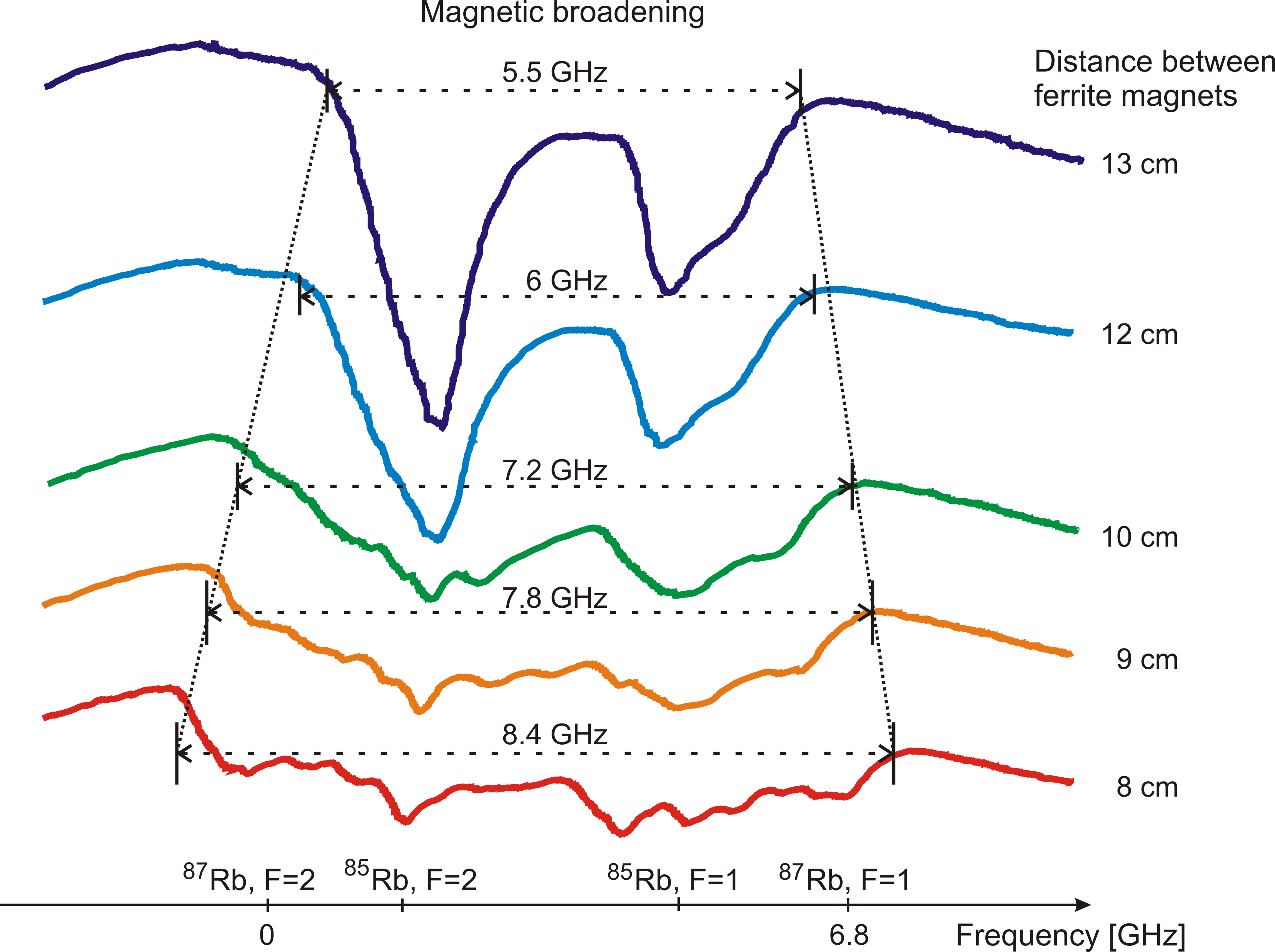}
\centering
\protect\caption{Large broadening of the lines of absorption filter due to transverse
magnetic field, tuned by changing distance between ferrite magnets.
For visualization purposes presented transmission spectra were measured
below operation temperature, at $T=60^{\circ}$ C. For our operation
we selected a distance of 9 cm corresponding to $10^{-2}$ T inside
the cell with $^{85}$Rb. Spectra are juxtaposed vertically, with
proportions preserved. \label{fig:MagneticBroadening}}

\end{figure}

\begin{figure}
 \includegraphics[width=1\columnwidth]{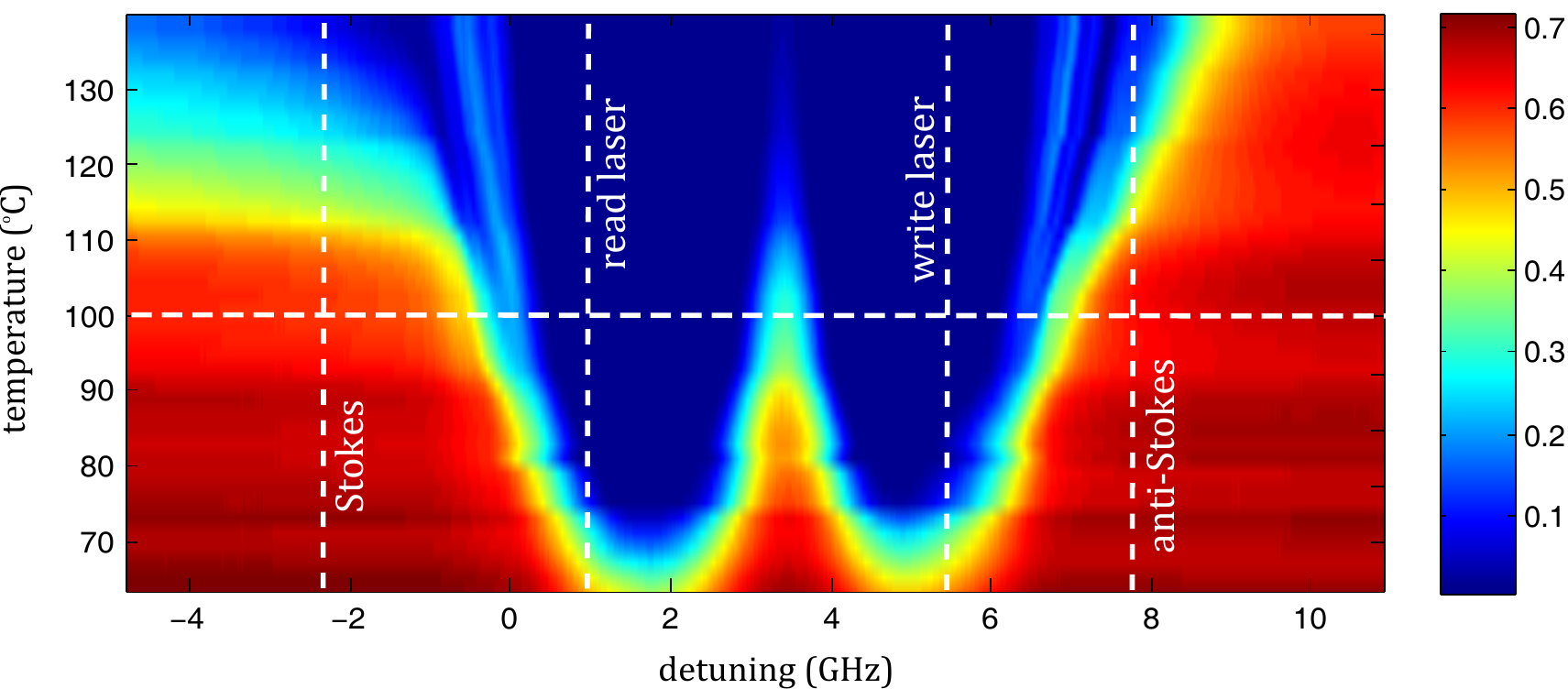}
 \centering
\protect\caption{$^{85}$Rb filter transmission spectra for a range of temperatures.
Absorption lines are significantly broadened due to transverse magnetic
field of approximately $10^{-2}$ T. The detuning is measured from
$F=2\rightarrow F'=2$ transition on the D1-line of $^{87}$Rb.}
\label{fig:absflt_temp} 
\end{figure}

The 30-cm long rubidium $^{85}$Rb cell in the transverse static magnetic
field, drawn in Fig. \ref{fig:TechnicalDrawingFilter}, is used to
stop the strong laser beams and transmit the Raman-scattered single
photons which are very narrowband but their frequencies lie out of
the wings region of the spectrum. The transverse magnetic field inside
the filter is emitted by the ferrite magnets located on both sides
of the rubidium cell. By changing the distance between the magnets,
we effectively change the magnetic field inside the cell and thence
the line broadening. The width of the absorption spectrum controlled
by the magnetic field ranges from 5.5 GHz up to 8.4 GHz, depending
on the distance between the magnets, as depicted in Fig. \ref{fig:MagneticBroadening}.
For our purposes we set the magnetic field to $10^{-2}$ T for a 9-cm
distance between magnets, to move $^{85}$Rb lines close enough to
$^{87}$Rb where driving lasers operate. 

The opaqueness and to some extent the width of the filter lines are
tuned by changing the temperature of the filter. The filter is heated
in a range of $20^{\circ}-140^{\circ}$ C to vary the optical density of atomic
vapor that grows exponentially with temperature. In practice the operational
range is limited, as for temperatures below $90^{\circ}$ C the absorption
of drive lasers is too low while above $120^{\circ}$ C the transmission
of the Raman-scattered light drops abruptly. 

The whole filter is inserted inside the two-layer soft-steel magnetic
shielding that confines the magnetic field inside the filter suppressing
the stray external fields down to $B_{ext}\simeq10^{-4}$ T. We heat
the vapors in the Rubidium cell using the high-power resistor mounted
under the cell which dissipates around $P\simeq80$ W power. 

Measured transmission spectra for different temperatures in the frequency
range of both lasers and the scattered light are shown in Fig. \ref{fig:absflt_temp}.
For large positive detunings the wing of the spectrum transmission
is smaller than for negative detunings due to the residual content
of $^{87}$Rb in the filter cell. For optimal temperature $T_{abs}=100^{\circ}$ C the suppression of the drive lasers is 50 dB while the transmission
of the Raman-scattered light is at the level of 80\%. The frequencies
of Stokes and anti-Stokes scattered light in Fig. \ref{fig:absflt_temp}
correspond to detunings $\Delta_{S}=-2.3$ GHz and $\Delta_{AS}=7.8$
GHz, respectively.

\subsection{Faraday filter}

To filter out the broadband fluorescence and four-wave mixing accompanying
write-in and readout from the memory, we utilize a Faraday filter
which consists of two orthogonal polarizers and a Faraday rotator
in between. In this arrangement the filter is opaque in the absence
of magnetic field. We use the standard configuration of the filter
\cite{Zielinska2012a,Kiefer2014}, but for temperature stabilization
purposes we apply water cooling to compensate the relatively high
power emitted by coils producing magnetic field. 

The rotation of the polarization occurs in a 30-cm long cell containing
the $^{87}$Rb isotope heated up to $100^{\circ}$ C for us to obtain appropriate
optical density of vapors. The magnetic field inside the filter is
about $B=10^{-2}$ T and directed along the beam. The average magnetic
field does not depend on the transverse position inside the cell,
i.e. it is near perfectly uniform in each plane perpendicular to the
laser beam direction. The imperfections of the magnetic field in the
transverse directions are smaller than $10^{-5}$ T. The cell is placed
in a PVC tube and heated with a current flowing through the coils
wound around the $^{87}$Rb glass cell. Other coils outside the PVC
tube produce strong magnetic field along the filter axis. The filter
is inside a soft-steel pipe with lids to contain the magnetic field
inside the filter. The whole system is additionally sealed in another,
outer pipe, where circulating water transfers out the heat emitted
by the coils.

The transmission through the system with two orthogonal polarizers
and the rotator between reads $t=t_{rot}\sin^{2}{\theta}$, where
$t_{rot}$ is the transmission through the rotator and $\theta$ is
the rotation angle. To elucidate the above expression, consider the
linearly polarized light entering the rotator decomposed in a left
(L) and right (R) circular polarization basis. Each of them has a
refractive index $n_{L,R}$ and an absorption coefficient $\alpha_{L,R}$.
Traveling through the rotator each circular polarization acquires
a phase $\exp(iK_{L,R}z)$, where $K_{L,R}=n_{L,R}\omega/c+i\alpha_{L,R}z$
is the complex wavevector. The resulting rotation angle due to the
difference between refractive indices equals $\theta=(n_{L}-n_{R})\omega/2c$.
For the sake of simplicity we assume equal absorption coefficients
for both circular polarization components $\alpha_{L}\simeq\alpha_{R}=\alpha$,
yielding the expression for the total transmission $t_{rot}=\exp(-\alpha z)$. 

\begin{figure}
 \includegraphics[width=1\columnwidth]{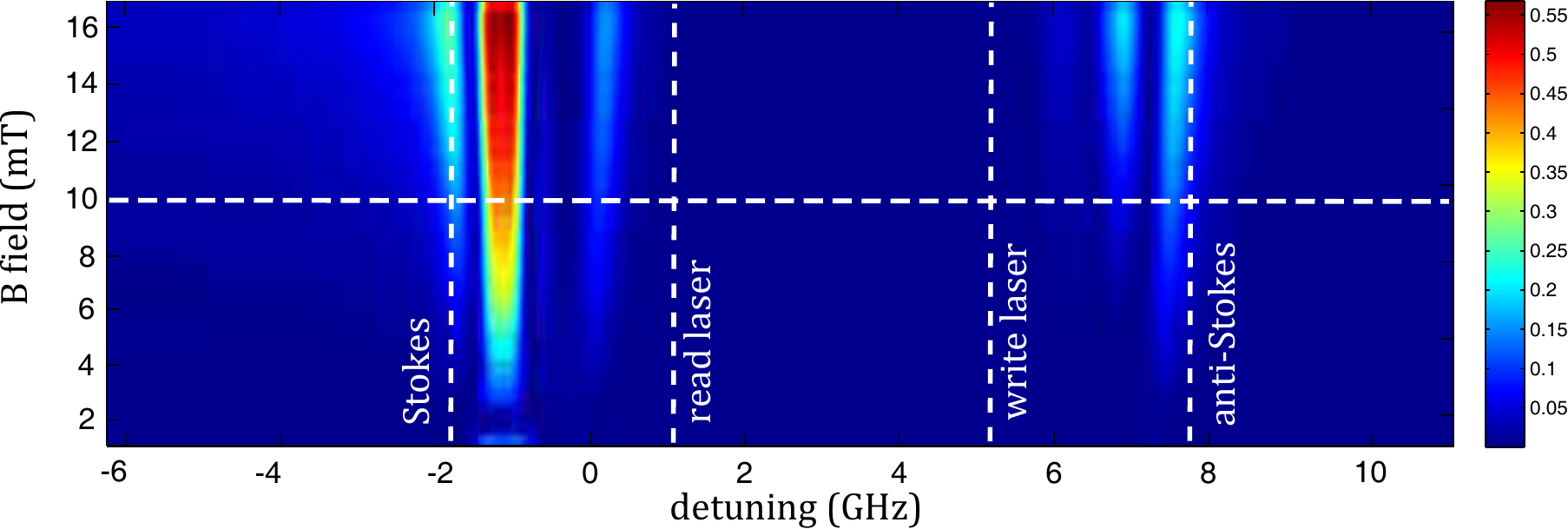}
 \centering
\protect\caption{Transmission spectra of the Faraday filter at $T=68^{\circ}$ C for different
values of magnetic field. The detuning is measured from $F=2\rightarrow F'=2$
transition on the D1 line of $^{87}$Rb.}
\label{fig:faraday_B} 
\end{figure}

The transmittance spectrum of the Faraday filter with two orthogonal
polarizers is presented in Fig. \ref{fig:faraday_B} for the whole
spectrum of rubidium-87 D1-line. The transmission spectrum is measured
for different values of the external magnetic field at the temperature
of $T=68^{\circ}$ C. For very weak magnetic fields there is almost no
transmission while in strong magnetic fields numerous separate high-transmission
bands emerge. They correspond to rotation angles $\theta=\pi+2\pi k$,
where $k$ stands for an integer number. 

\begin{figure}
 \subfigure[Part of Rb spectrum in the region of F=1 ground-state sublevel transition.]{
\includegraphics[width=1\columnwidth]{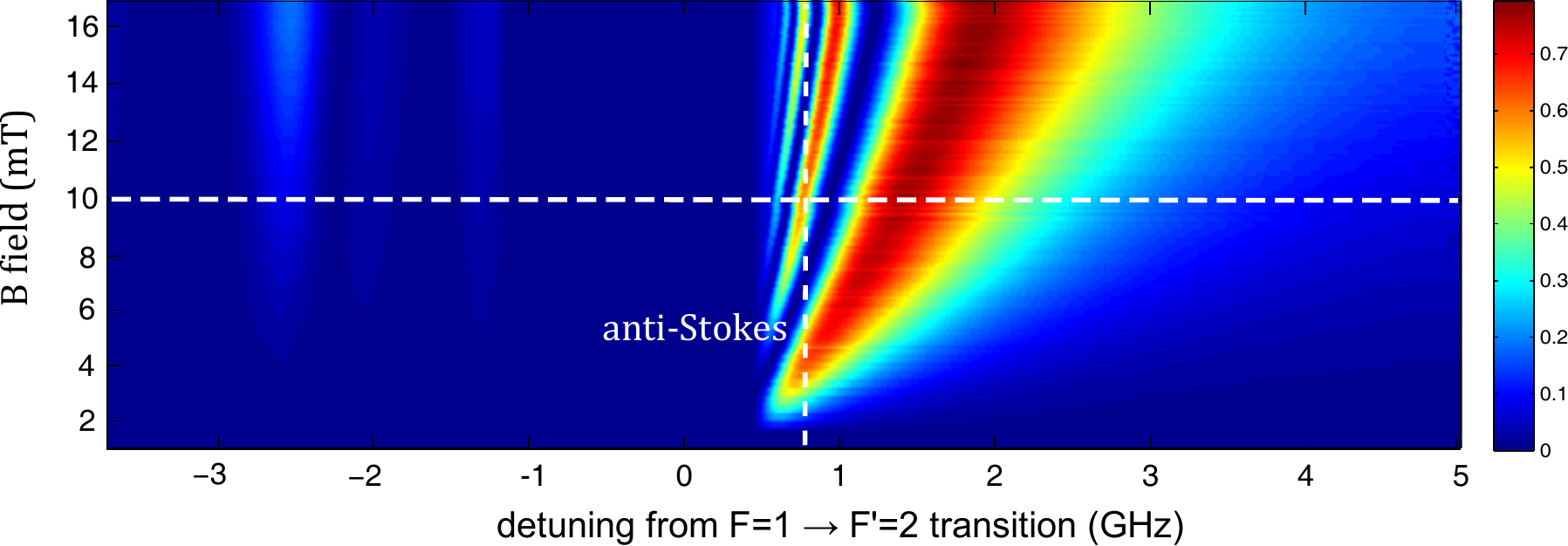}}\hspace{6pt} \centering
\\
\subfigure[Part of Rb spectrum in the region of F=2 ground-state sublevel transition.]{
\includegraphics[width=1\columnwidth]{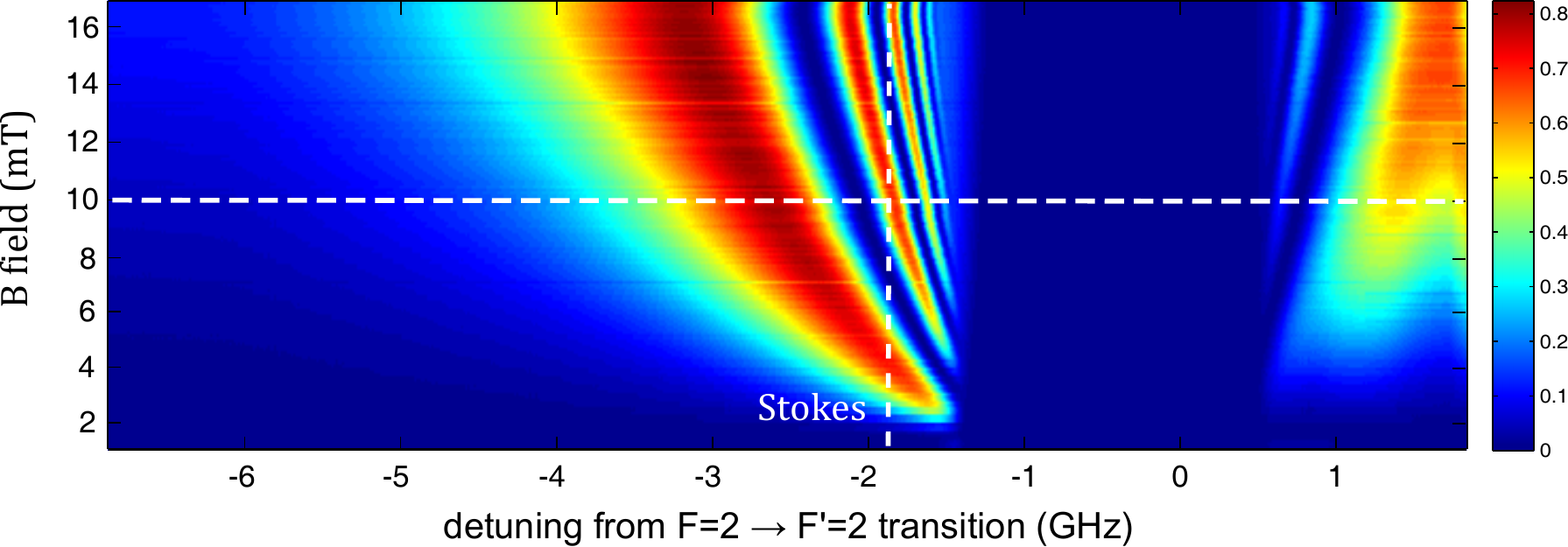}}\centering
\protect\caption{Transmission spectra of the Faraday filter with a hot cell at $T=110^{\circ}$
C for different values of magnetic field zoomed to the vicinity of
the $^{87}$Rb resonances.}
\label{fig:faraday_zoom_hot} 
\end{figure}

\begin{figure}
 \includegraphics[width=1\columnwidth]{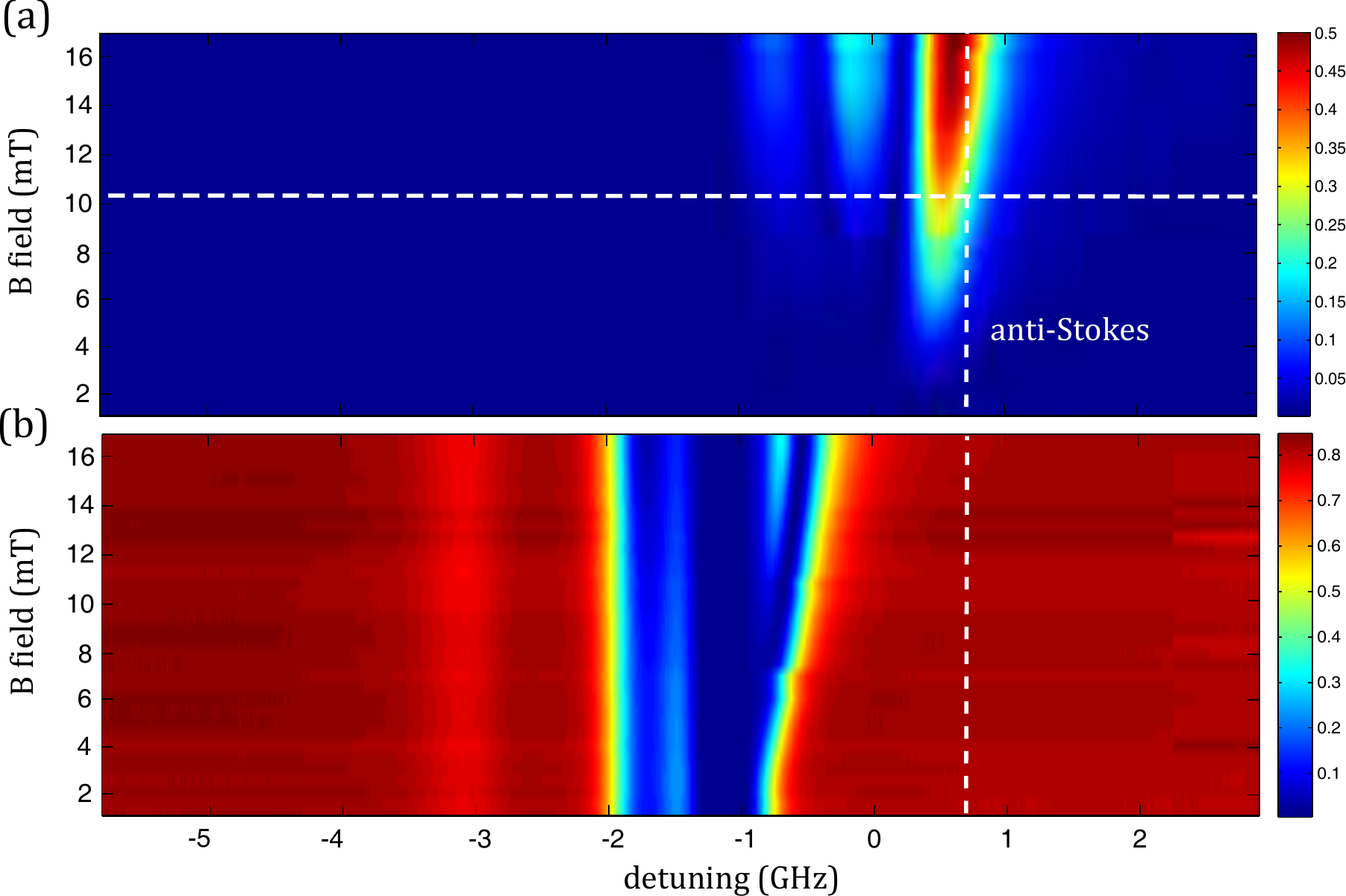}
 \centering
\protect\caption{Transmission spectra of the Faraday filter at $T=90^{\circ}$ C with (a)
perpendicular and (b) parallel polarizers for reference. The detuning
is measured from $F=1\rightarrow F'=1$ transition on the D1 line
of $^{87}$Rb. }
\label{fig:faraday_zoom_polariz} 
\end{figure}

Fig. \ref{fig:faraday_zoom_hot} presents analogous spectra but measured
in a detailed frequency scale at a higher temperature $T=110^{\circ}$
C, separately for the two different regions of the D1 line of Rubidium:
near the the $F=1\rightarrow F'=2$ and $F=2\rightarrow F'=2$ transitions,
respectively. Fig. \ref{fig:faraday_zoom_polariz} compares two relative
orientations of polarizers parallel and perpendicular to each other
at $T=90^{\circ}$ C.

For the perpendicular polarizers the transmission $t_{\bot}=t_{rot}\sin^{2}{\theta}$
reaches its maximum for a rotation angle $\theta=\pi+2\pi k$ provided
the transmission through rotator $t_{rot}$ is nonzero. Although in
the vicinity of the resonance the rotation may be high, the rotator
becomes opaque \cite{Zielinska2012a}. On the other hand, further
away from the resonances the net polarization rotation is very small,
since both circular polarization components are subject to virtually
the same retardation. Increase in the magnetic field pushes the transmission
line further from the resonances. Practically the filter transmits
frequencies detuned by 1-3 GHz from any of the four resonances of
$^{87}$Rb D1 line, depicted in Fig. \ref{fig:sample_transmissions}
(a). 

\begin{figure}
 \includegraphics[width=1\columnwidth]{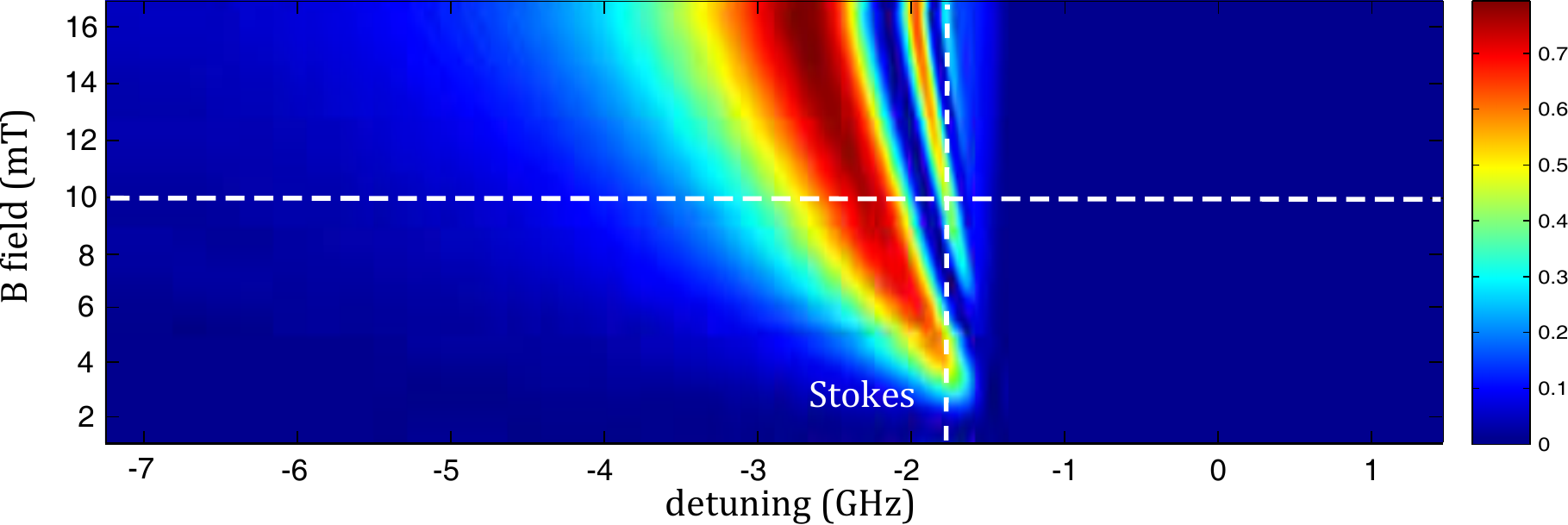}\centering
\protect\caption{Transmission spectra of both filters for different magnetic fields
inside the Faraday filter. The detuning is measured from $F=2\rightarrow F'=2$
transition on the D1-line of $^{87}$Rb. Temperatures of the filters
were $T_{\mathrm{abs}}=100^{\circ}$ C and $T_{\mathrm{Farad}}=102^{\circ}$
C, respectively for the absorption and Faraday filer.}
\label{fig:trasm_map_both} 
\centering
\end{figure}

\subsection{Transmission of combined filters}

The transmission through both filters for the D1 line Rubidium spectrum
is depicted in Fig. \ref{fig:trasm_map_both}. Both filters transmit
light at the frequencies detuned to the red from the $F=2\rightarrow F'=2$
resonant transition. The frequency corresponding to the maximum transmission
can be adjusted by proper settings of the Faraday filter's temperature
and the internal magnetic field. For positive detunings we observe
a sharp edge of absorption region resulting from the presence of the
absorption filter.

We found the optimal conditions for the operation of the filtering
system to be: $T_{\mathrm{abs}}=100^{\circ}$ C and $T_{\mathrm{Farad}}=102^{\circ}$
C, with the magnetic field amplitude inside the Faraday filter $B=10^{-2}$
T. For such settings the combined transmission through absorption
and Faraday filters is depicted in Fig. \ref{fig:sample_transmissions}
Remarkably for Raman-scattered light detuned from the $F=2\rightarrow F'=2$
resonant transition by $\Delta_{S}=-2.3$ GHz and $\Delta_{AS}=7.8$
GHz we achieved a high transmission of 65\% and 40\% for Stokes and
anti-Stokes light respectively.

\section{Performance - filtering photons from the memory}

To qualitatively assess the performance of the filtering system, we
measured Stokes photons along with subsequently retrieved anti-Stokes
photons in a similar manner as \cite{Chrapkiewicz2012}, but in a
low gain regime exciting only a few photons per spatial mode \cite{Chrapkiewicz2015}.
We observe an increase in the signal-to-noise ratio owing to the filtering
system by measuring average intensities and correlations between the
number of photons generated in the write-in and readout process.

\begin{figure}[t]
 \includegraphics[width=1\columnwidth]{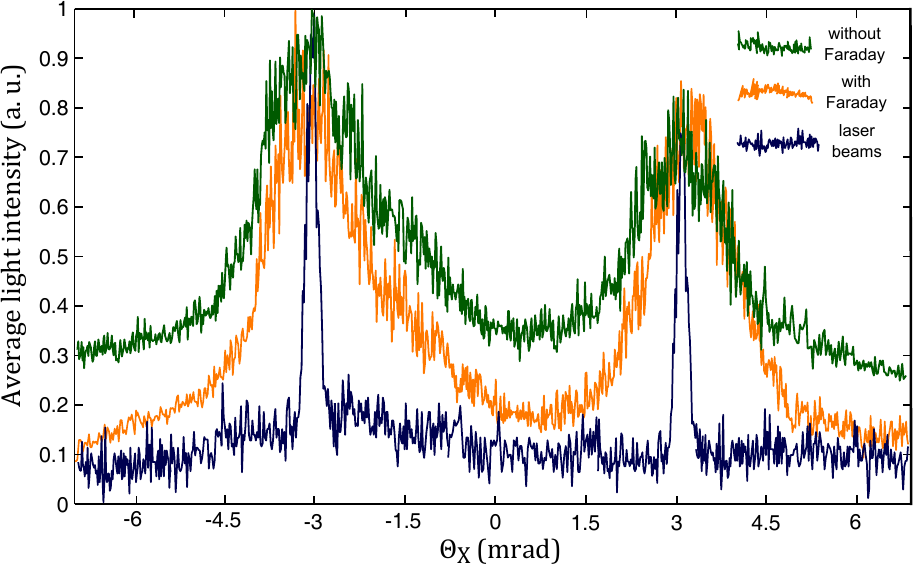} \centering 
 \protect\caption{Normalized average light intensities of angularly broad Stokes (left)
and anti-Stokes (right) scattering registered by the I-sCMOS camera
along the $\theta_{X}$ axis. Application of the Faraday filter considerably
diminishes the isotropic fluorescence background. Raman scattering
is compared with angularly narrow laser beam leakage which dominates
without absorption filter. }
\label{fig:result_section} 
\end{figure}

The average intensities from $2\times10^{5}$ frames registered on
the I-sCMOS camera are depicted in Fig. \ref{fig:result_section}.
The first line presents the angularly broad Stokes and anti-Stokes
light intensities without the Faraday filter, while the second one
includes Faraday filter operation where the isotropic noise background
is greatly suppressed. For comparison the residual laser beams (angularly
narrow) leakage is let through the filters in the system by with cold $^{85}$Rb
filter, as depicted in the third line. 

\begin{figure}[b]
 \includegraphics[width=1\columnwidth]{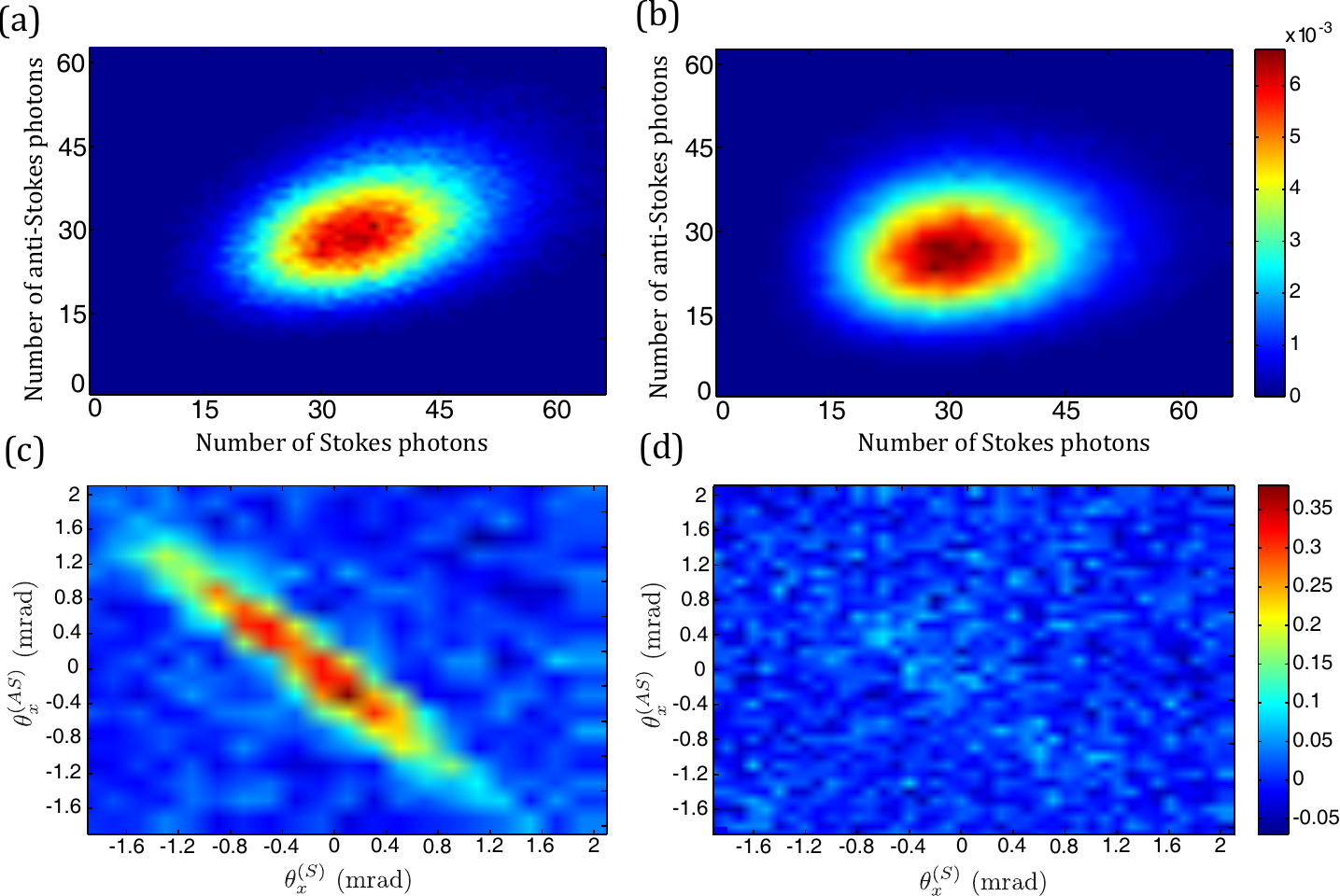}
 \centering
\protect\caption{Comparison of measured joint photon statistics (a-b) and spatial correlations
between the Stokes and anti-Stokes photons (c-d) with (left panel)
and without (right panel) Faraday filter. The joint count statistics
of centrally emitted Stokes and anti-Stokes (top row) $p(n_{S},n_{AS})$
and maps of intensity correlations between different directions (bottom
row).}
\label{fig:result_maps} 
\end{figure}

The joint statistics $p(n_{S},n_{AS})$ of counts between Stokes and
anti-Stokes photons from the write-in and readout processes are depicted
in Fig. \ref{fig:result_maps} (a) and (b), with and without the Faraday
filter, respectively. We count the number of Stokes photons $n_{S}$
scattered on the camera region around the write-in laser beam (6 mrad
in diameter) and the number of anti-Stokes photons $n_{AS}$ around
the readout beam.

Comparing the maps of the joint statistics, one can see that application
of the Faraday filter enables observation of a correlation between
the amount of Stokes and anti-Stokes photons scattered in each realization
of the experiment. The ratio of the size of the diagonal and anti-diagonal
joint probability distribution is a measure of correlation coefficient
and thus the signal-to-noise ratio in the registered data. Without
the Faraday filter the number of anti-Stokes scattered photons is
almost independent of the number of Stokes-scattered photons. This
is due to spurious incoherent light which is intense enough to cover
the Raman scattering signal. 

To demonstrate the multimode capacity of our memory and filter system
we conduct direction-resolved measurements of correlations. To this
end we divide the camera region along the horizontal line to the left
and right around the write and read beams into small circular regions
$A_{i}$ and $A_{j}$, each of the area of 0.02 mrad$^{2}$ and with
0.7-0.8 single photon on average. We calculated the correlation coefficient
$C_{ij}=\langle\Delta n_{S}(A_{i})\Delta n_{AS}(A_{j})\rangle/\sqrt{\langle(\Delta n_{S}(A_{i}))^{2}\rangle\langle(\Delta n_{AS}(A_{j}))^{2}\rangle}$
between the number of Stokes $n_{S}(A_{i})$ and anti-Stokes $n_{AS}(A_{j})$
photons scattered in the circular regions $A_{i},A_{j}$ around the
write and read beam, respectively. The maps of correlation coefficient
between number of the Stokes and anti-Stokes photons scattered into
each pair of circular regions are depicted in Figs. \ref{fig:result_maps}(c)
and (d). Using Faraday filter the four-wave-mixing contribution \cite{Dabrowski2014}
has been eliminated.

Then we compare the number of coincidences of Stokes and anti-Stokes
photons from each of these small regions. Including the Faraday filter
we are able to observe the elliptic structure which is evidence for
generation of correlated photon pairs with a correlation coefficient
of 0.38 which is smaller than unity because of the fluorescence signal
from the quantum memory cell. 

Without the narrowband Faraday filter we do not observe any correlations
on the illumination level of $\bar{n}=30$ photons per camera frame
(from which $\bar{n}_{b}=1.5$ is the light-independent equivalent
background illumination noise of the image intensifier). The correlation
coefficient between the Stokes and anti-Stokes photons scattered in
the opposite directions extracted from the measured data is as little
as ca. 0.03 on average which is negligible as compared to Fig. \ref{fig:result_maps}(c)
with Faraday filter. Moreover, in Fig. \ref{fig:result_maps}(d) there
is no visible structure, which is due to the large amount of fluorescence
light comparable to the Raman-scattered photons which are correlated.

\section{Conclusions}

In conclusion we built and characterized a two-step filtering setup
which enables observation of the correlations between Stokes and anti-Stokes
Raman scattering at a single-photon-level. 

Polarization pre-filtering and a rubidium absorption filter are used
to eliminate the laser beams leakage by attenuating them by more than
$10^{9}$, while an additional combination with a Faraday filter yields
a total attenuation of $10^{11}$.

Moreover the Faraday filter blocks the fluorescence and the four-wave-mixing
noise coming from the atomic memory medium. The filter transmits only
the narrowband spectrum of frequencies corresponding to the Raman-scattered
light in $^{87}$Rb. In the case of our filter construction it corresponds
to the frequencies detuned approximately by $\Delta_{S}=-2$ GHz and
$\Delta_{AS}=7$ GHz from the $F=2\rightarrow F'=2$ resonant transition
for Stokes and anti-Stokes light, respectively. Noticeably, regardless
of such small detunings we achieve a transmission for the Raman-scattered
light as high as 50-60\%, depending on the Faraday filter settings.

Our approach is robust and efficient. Which is more, the frequencies
with high transmission can be tuned by several hundreds of MHz by
changing the magnetic field amplitude inside the Faraday filter, whereas
absorption edges of $^{85}$Rb filter can be moved magnetically as
widely as up to 8.4 GHz. Eventually we can obtain the optimal, high
transmission on the Raman-scattered light frequencies while blocking
frequencies corresponding to the broadband fluorescence, laser beams
and four-wave-mixing.

The performance of the system in the regime of a few atomic collective
excitations is evaluated in statistical measurements of the scattered
photons. The obtained correlation maps are strong evidence for the
directional coincidences between Raman-scattered light produced in
the write-in and readout processes of quantum memory. However, the output signal is dominated by
noise and no correlation appears above the background if no filtering is applied.

Nowadays, many groups are utilizing warm atomic and molecular gases
for quantum memories \cite{Bashkansky2012,Eisaman2005,Jiang2004,Michelberger2014a,VanderWal2003,Bustard2015,Bussieres2013,Bustard2013,Chrapkiewicz2012,DeAlmeida2015,Eisaman2004a,Glorieux2012,Heifetz2004,Hosseini2009,Hosseini2011,Lvovsky2009,Manz2007,Nunn2008,Sprague2013,Sprague2014}.
Application of the filtering system we present can enhance the fidelity
of retrieved photons and extend other teams' operational regime into
the multimode quantum memory \cite{Nunn2008,Surmacz2008}. Undoubtedly
such system will be essential to measure non-classical $g^{(2)}$
cross-correlation function between stored and retrieved photons \cite{Bashkansky2012,Lee2011,Kasperczyk2015},
which is prerequisite for further applications in quantum information
processing. Importantly presented filtering solution can find applications
far beyond quantum memories including Raman spectroscopy \cite{Lin2014}
and biological applications \cite{Uhland2015}.

\section*{Acknowledgments}

We acknowledge discussions with Micha\l{} Parniak. Czes\l{}aw Radzewicz
and Cezary Samoj\l{}owicz, generous support of Konrad Banaszek, as
well as technical support of Jaros\l{}aw Iwaszkiewicz, Marcin Piasecki
and Szymon \.Zuchowski.

\section*{Funding}

This project was financed by the National Science Centre No. DEC-2011/03/D/ST2/01941
and DEC-2013/09/N/ST2/ 02229 and PhoQuS@UW (Grant Agreement no. 316244)
co-financed by the Polish Ministry of Science and Higher Education.
R.C. was supported by Foundation for Polish Science (FNP).

\end{document}